\newcommand{\Gup}{\overset{\sim}{G}}
\newcommand{\Glow}{\underset{\sim}{G}}

\newcommand{\GDF}{\overset{\sim}{\mathcal{G}}}

\newcommand{\deltaABDE}{\delta_{\nu_{n+1}  \ldots \nu_{2n}}^{\nu_1  \ldots \nu_n}}

\documentclass[aps,floats,twocolumn,prb,showpacs]{revtex4-1}
\usepackage{graphicx}
\usepackage{color}
\usepackage{upgreek}
\usepackage{setspace}
\usepackage{amssymb}
\usepackage{amsmath}
\usepackage{pstricks}
\usepackage{dsfont}
\usepackage{fancyhdr}
\usepackage{array}
\usepackage{multirow}
\usepackage{placeins}
\usepackage{todonotes}

\begin{document}
\title{Local correlation functions of arbitrary order for the Falicov-Kimball model}
\author{ T. Ribic${}^1$, G. Rohringer${}^{1,2}$, and K. Held${}^1$}
\affiliation{${}^1$Institute of Solid State Physics, TU Wien, 1040
Vienna, Austria ,${}^2$Russian Quantum Center, Novaya street, 100, Skolkovo, Moscow region 143025, Russia}
\date{Version 0.95, \today}

\begin{abstract}
Local $n$-particle vertex functions represent the crucial ingredient for diagrammatic extensions of dynamical mean field theory (DMFT). Hitherto their application has been restricted --with a few exceptions-- to the  $n\!=\!2$-particle vertex while higher-order vertices have been neglected. In this paper we derive a general analytical expression for the local $n$-particle (one-particle reducible) vertex of the Falicov-Kimball model (FKM). We observe that the magnitude  of such vertex functions itself strongly increases with the number of particles $n$. On the other hand, their effect on generic Feynman diagrams remains rather moderate due to the damping effect of the Green's functions present in such diagrams. Nevertheless, they yield important contributions to the self-energy corrections calculated in diagrammatic extensions of DMFT as we explicitly demonstrate using the example of dual fermion (DF) calculations for the two-dimensional FKM at quarter filling of the stationary $f$-electrons. Here corrections to the self-energy obtained from the three-particle vertex are indeed comparable in magnitude to corresponding corrections stemming from the two-particle vertex.

 \end{abstract}

\maketitle


\section{Introduction}
\label{Sec:Intro}
 With the establishment of dynamical mean field theory (DMFT) \cite{Metzner89a,MuellerHartmann89a,Georges92a,Jarrell92a,Georges96a} as a state-of-the-art method for calculating strongly correlated electron models \cite{Georges96a} and materials \cite{dmft1,dmft2}, the scientific frontier moved on to extensions of DMFT that include the important local correlations of DMFT but also non-local correlations beyond. One route to this end are cluster extensions where a finite cluster of sites is surrounded by a DMFT bath \cite{DCA,clusterDMFT,LichtensteinDCA,Maier04}.
More recently, diagrammatic extensions have been proposed as a vivid alternative \cite{DGA1,Kusunose06,Katanin09,DualFermion,DGAintro}. These have in particular the advantage that long-range and short-range correlations beyond local ones of DMFT are treated on an equal footing, and that realistic materials calculations \cite{AbinitioDGA} are possible. Note that the numerical effort of cluster extensions generally grows exponentially with the number of sites times the number of orbitals, which severely restricts their application.

Several closely related  diagrammatic extensions of DMFT have been proposed: the dynamical vertex approximation (D$\Gamma$A) \cite{DGA1,Kusunose06,Katanin09}, the dual fermion (DF) \cite{DualFermion}, the dual boson\cite{DualBoson} and the non-local expansion scheme \cite{Li15}, the one-particle irreducible approach (1PI) \cite{1PI}, the merger of DMFT with the functional renormalization group (DMF$^2$RG)  \cite{DMF2RG},  the  triply- and quadruply-irreducible local expansion \cite{Ayral15,Ayral16} and DMFT+fluctuation exchange (FLEX) \cite{Kitatani15}. All these approaches start from the local, but fully frequency dependent vertex, and diagrammatically construct non-local correlations beyond DMFT from it. Particular highlights achieved by applying these novel approaches are the calculation of the critical \cite{Rohringer11,Hirschmeier15}  and quantum critical \cite{Schaefer17} exponents of the Hubbard model  and those of the Falicov Kimball model \cite{Antipov14}, as well as establishing the insulating nature of the paramagnetic phase  in the half-filled Hubbard model on a square lattice at  arbitrarily small interactions \cite{Schaefer15}.

The DF and 1PI approach constitute a systematic expansion of the self-energy of the system in local one-, two- and more-particle correlation (vertex) functions which is, in principle, exact when including {\sl all} $n$-particle vertices as well as all diagrams built from these. In practice, these approaches are  truncated at the two-particle vertex level which constitutes the most fundamental approximation regarding these methods \cite{footnote1}. 

The main reason for this truncation is that the complexity and numerical effort is already at the edge of what is possible for the $n=2$ particle vertex, for which  the full dependence on three frequencies needs to be taken into account \cite{vertex}. At least this is true for the Hubbard model within DMFT, where arguably the most efficient way to calculate the corresponding local two-particle vertex is by means of quantum Monte-Carlo simulations \cite{Rubtsov04,Werner06,Gull11} with worm sampling \cite{Gunacker15} and improved estimators \cite{Gunacker16}. While individual contributions of the local three-particle vertex of the DMFT Hubbard or impurity model have been calculated this way \cite{Gunacker16}, obtaining its full dependence on five frequencies exceeds presently available resources regarding both, computational time and memory. From this two-particle vertex Feynman diagrams are constructed and yield  non-local correlations beyond DMFT.  


The Falicov Kimball model (FKM) \cite{Falicov69} has a much simpler structure than the Hubbard model. In contrast to the latter it describes only one species of  itinerant electrons that interact with  the other, localized species. Because of this localization, one can consider the FKM as a model for (annealed) disorder instead of interaction. Because of its simpler structure the FKM has a long tradition of (semi-)analytical calculations that are not possible for the Hubbard model. The FKM has been commonly utilized as a test-bed for new approaches and concepts. Indeed the FKM can be solved analytically in DMFT \cite{Brandt89,Freericks03},
including charge density wave ordering \cite{vanDongen90} and transport properties \cite{Freericks00}. Vertex corrections to the conductivity\cite{Janis10,Pokorny13} and Anderson localization\cite{Antipov16} have been discussed; and it has been studied in DF \cite{Antipov14} and 1PI \cite{Ribic16}.
 
The analytical calculation of the local two-particle vertex of the FKM\cite{Freericks03,Shvaika,Ribic16} shows a
much simpler, reduced frequency structure in comparison to the Hubbard model \cite{vertex}. That is, the two-particle vertex  has only two contributions since no energy can be transferred to the localized electrons:
 (i) an  $\omega\neq0$ ($\omega$: transferred bosonic frequency) contribution
where the two fermionic frequencies have to be the same $\nu=\nu'$ \cite{Freericks03} and (ii)  an  $\omega=0$ contribution \cite{Shvaika,Ribic16}.

In this paper, we analytically calculate the full $n$-particle  vertex of the FKM for arbitrary $n$ and analyze its magnitude as a function of $n$. This is a very relevant issue since the typical truncation of the self-energy diagrams in the DF and 1PI theory at the two-particle vertex level relies on the ``smallness'' of higher-order vertex functions. We  exemplary investigate the importance of such higher-order vertices in the DF theory by including the $n=3$ particle vertex in the DF correction to the DMFT self-energy. While at half-filling this three-particle vertex vanishes, it yields a sizable contribution to the nonlocal DF correction of the DMFT self-energy in  the case of quarter filling, indicating that in a general situation some caution has to be taken when cutting the DF expansion at the two-particle vertex level.

The outline of our paper is the following: In Section \ref{Sec:nPartVert}, we derive the analytical formula for the $n$-particle connected Green's and vertex functions of the FKM within DMFT. To this end, we first calculate in Section \ref{sec:calcdmftgreen} the  $n$-particle Green's function, from which we remove all disconnected contributions in Secs.~\ref{sec:decomp1} and \ref{sec:decomp2}. In Section \ref{sec:propC} we analyze how the prefactor of the $n$-particle vertex grows with $n$ and provide an analytic estimate for its asymptotic behavior. From this we deduce in Sec.~\ref{sec:ordermagnitude} the magnitude of the contribution of such vertices in generic self-energy diagrams of DF and 1PI. In Section  \ref{sec:numerical} we then calculate some selected correction terms for the DF self-energy stemming from the $n=3$ particle vertex and compare these to the usual corrections from two-particle vertex functions only. Finally, Section  \ref{sec:conclusion} is devoted to a summary of our main results and an outlook.

\section{Analytic derivation of the $n$-particle vertex in the FKM}
\label{Sec:nPartVert}

The Hamiltonian of the spin-less (one-band) Falicov Kimball model reads
\begin{equation}
 \label{equ:defhamilt}
 \hat{\mathcal{H}}=-t\sum_{\langle ij \rangle}\hat{c}^{\dagger}_i \hat{c}_j + U \sum_i\hat{c}^{\dagger}_i\hat{c}_i\hat{f}^{\dagger}_i\hat{f}_i -\mu_c\sum_i\hat{c}^{\dagger}_i\hat{c}_i-\mu_f\sum_i\hat{f}^{\dagger}_i\hat{f}_i
\end{equation}
where $\hat{c}^{(\dagger)}_i$ annihilates (creates) an itinerant $c$-electron and, correspondingly, $\hat{f}^{(\dagger)}_i$ annihilates (creates) a localized $f$-electron at the lattice site $i$; $t$ represents the hopping amplitude for a mobile electron between nearest-neighbor sites $\langle ij\rangle$. The two electron species interact with each other via a purely local (Hubbard-like) interaction parametrized by the interaction strength $U$; $\mu_c$ and $\mu_f$ denote the chemical potentials for the itinerant and localized electrons, respectively, which determine the density of the corresponding particles. Throughout the paper all energies will be measured in terms of $4t$ representing the half band-width for the FKM on a $2d$ square lattice;  $T=1/\beta$ is the temperature.

In the following we will consider the DMFT solution of model (\ref{equ:defhamilt}) and derive analytical expressions for the corresponding local (connected) $n$-particle DMFT vertex function for arbitrary $n$. To this end we systematically subtract all disconnected contributions from the full $n$-particle Green's function $G^{(n)}$. As we will show, this is achieved by a two-step procedure where (i) all diagrams which can be written as a product of a one-particle Green's function $G^{(1)}$ and a remainder are eliminated from $G^{(n)}$. Subsequently we (ii) remove in the second step all disconnected products among purely higher-particle Green's functions ($G^{(2)},G^{(3)},\ldots$). This yields --after amputating the outer legs via dividing by the corresponding one-particle Green's functions $G^{(1)}$-- the final result for the $n$-particle local DMFT vertex function for the FKM.

Let us stress that all results and derivations presented in this section are independent of the underlying lattice type (e.g., square, cubic, Bethe, etc.). The density of itinerant and localized electrons only enters through the local one-particle Green's function. 

\subsection{Calculation of DMFT $n$-particle Green's functions}
\label{sec:calcdmftgreen}

Let us start our considerations by recalling that the (local) single particle impurity Green's function for $c$-electrons of the FKM in the framework of DMFT is just given as\cite{Freericks03}
\begin{equation}
 \label{equ:defonepartgreen}
 G(\nu)=p_1\underset{\equiv\Gup(\nu)}{\underbrace{\frac{1}{i\nu+\mu_c-\Delta(\nu)-U}}}+p_2\underset{\equiv\Glow(\nu)}{\underbrace{\frac{1}{i\nu+\mu_c-\Delta(\nu)}}},
\end{equation}
where $\nu=\pi T(2n_\nu+1)$, $n_\nu\in\mathds{Z}$, is a fermionic Matsubara frequency and $\Delta(\nu)$ denotes the hybridization function of DMFT, encoding the dispersion relation of the system and uniquely defining the local DMFT problem. $\Glow(\nu)$ represents the (non-interacting) single-particle Green's function for the itinerant $c$-electrons if no localized $f$-electron is present at the impurity site; $\Gup(\nu)$ corresponds to the same (non-interacting) Green's function but in the presence of a localized electron which, however, just increases the energy level for the itinerant one by $U$. Here, the weight $p_1\!=\!\langle\hat{f}^{\dagger}\hat{f}\rangle$ corresponds to the density of localized electrons at each lattice site (i.e., to the probability for finding such an electron at the given site) and $p_2=1-p_1$. In other words, the local DMFT single-particle Green's function for the itinerant electrons is given just by the weighted average of two non-interacting Green's functions that correspond to the presence or absence of a localized electron on the DMFT impurity. The localized electron merely acts as a scattering potential for the itinerant particles.

We start our derivation by calculating the  $n$-particle Green's function
on the DMFT impurity, and recall that it is defined as
\begin{multline}
G^{(n)} ( \tau_1 , \tau_ 2 , \ldots,\tau_{2n} ) = (-1)^{n} \\ \langle \textrm{T} \left[ \hat{c} (\tau _{1}) \hat{c}^{\dagger} (\tau _{2}) \hat{c}(\tau _{3}) \hat{c}^{\dagger} (\tau _{4}) \ldots \hat{c}(\tau _{2n-1}) \hat{c}^{\dagger}(\tau _{2n}) \right] \rangle .
\label{Gt}
\end{multline}
Here,  $\langle \,\ldots \rangle$ denotes the thermal average, and T is the Wick operator ordering the (imaginary) times $\tau_1 \ldots \tau_{2n}$.  For the Fourier transform, we chose the following convention --for convenience in later steps of the derivation\cite{note_fourier}:
\begin{multline}
G^{(n)} ( \nu_1 , \nu_ 2 , \ldots, \nu_n;  \nu_{n+1} , \nu_{n+2} , \ldots,\nu_{2n} ) = \\ \dfrac{1}{\beta^n} \int_0^\beta d\tau_1 \ldots \int_0^\beta  d\tau_{2n}  G^{(n)} ( \tau_1 , \tau_ 2 , \ldots ,\tau_{2n} ) \\ e^{-i (\nu_1 \tau_2 + \nu_2 \tau_4 + \ldots + \nu_{n} \tau_{2n} - \nu_{n+1} \tau_{1} \ldots - \nu_{2n} \tau_{2n-1})} .
\label{Gf}
\end{multline}
Because of the non-interacting-like nature of the FKM, we can determine the $n$-particle Green's functions quite straightforwardly.
To this end, we do not need to calculate the sum of all possible $n$-particle diagrams for the $c$-$f$ interaction $U$  directly. Instead we can express the local $G^{(n)}$  as an average of two non-interacting $n$-particle Green's functions, corresponding to the presence or absence of an $f$-electron which acts as a local disorder potential. As we have two non-interacting Green's functions, we  can simply apply Wick's theorem. This yields all possible products of one-particle Green's functions $\Glow$ or $\Gup$, with or without  a localized electron on the impurity, see Eq.~(\ref{equ:defonepartgreen}).
The resulting, averaged Green's function
\begin{multline}
G^{(n)} ( \nu_1 , \nu_ 2 , \ldots, \nu_n;  \nu_{n+1} , \nu_{n+2} , \ldots,\nu_{2n} ) = \\ \deltaABDE \left[ p_1 \Big( \Gup (\nu_1) \Gup (\nu_2) \ldots\Gup (\nu_n) \Big) +\right.\\ \left.p_2 \Big( \Glow (\nu_{1}) \Glow (\nu_{2})\ldots \Glow (\nu_{n})\Big) \right]
\label{Gn}
\end{multline}
contains all connected as well as disconnected diagrams.
Here $\nu_1,\nu_2 \ldots \nu_n$ correspond to the ($n$) frequencies of the incoming and $\nu_{n+1},\nu_{n+2}\ldots \nu_{2n}$ to the ($n$) frequencies of the outgoing particles.
$\deltaABDE$ denotes a generalized Kronecker delta which guarantees energy (frequency) conservation: It is $1$ if the entering frequencies are an even permutation of the leaving ones, $-1$ if they are an odd permutation and $0$ otherwise. Moreover, it will be $0$ if an incoming/outgoing frequency appears more than once. Note that  $\deltaABDE$ generates  $n!$ summands that correspond to the $n!$ Wick contractions.

The following parts of the derivation of the $n$-particle vertex are technical in nature. Readers interested in the analytical expressions for and the impact of higher-order vertices, but not the technical details of their derivation can safely skip the rest of this section. The final result of our derivation is given in Eq.~(\ref{nVert}).

Eq. (\ref{Gn}) corresponds to the {\sl full} local DMFT $n$-particle Green's function and, hence, contains contributions from disconnected as well as from fully connected diagrams. To extract the latter --and from this the $n$-particle vertex-- we have to remove all disconnected contributions from Eq. (\ref{Gn}). At  first glance, the most direct way to achieve this seems to be  simply to subtract all terms of the form $G^{(m_1)}\ldots G^{(m_k)}$ with $\sum_{l=1}^k m_l=n$ for all possible partitions $m_l$ and all possible frequency assignments for a given partition. Such a procedure, however, would lead to a massive over-counting of subtraction terms as one can easily see from the following argument: Consider for instance the contributions $G^{(1)}G^{(n-1)}$ and $G^{(2)}G^{(n-2)}$ which certainly have to be subtracted from $G^{(n)}$. However, the first term $G^{(n-1)}$ includes a contribution $G^{(1)}G^{(n-2)}$ and the second term $G^{(2)}$ contains $G^{(1)}G^{(1)}$ so that it becomes obvious that the expression $G^{(1)}G^{(1)}G^{(n-2)}$ appears in {\sl both} subtraction terms. This indeed leads to the above mentioned over-counting problem when naively removing such contributions. 

In the following, we will, hence, take another path trying to avoid the above mentioned ``over-subtraction''. To this end, let us first consider the following algebraic identities for $G(\nu)$, $\Gup(\nu)$ and $\Glow(\nu)$:
\begin{eqnarray}
\Gup (\nu) &=& p_1 \Gup (\nu) + p_2 \Gup (\nu) + p_2 \Glow (\nu) - p_2 \Glow (\nu)\nonumber
\\ &=&G^{(1)} (\nu) + p_2 \big[ \Gup (\nu) - \Glow (\nu) \big]
\\ \Glow (\nu) &=& p_1 \Glow (\nu) + p_2 \Glow (\nu) + p_1 \Gup (\nu) - p_1 \Gup (\nu)  \nonumber
\\ &=&G^{(1)} (\nu) - p_1 \big[ \Gup (\nu) - \Glow (\nu) \big].
\end{eqnarray}
Substituting these relations into Eq. (\ref{Gn}) recasts the expression of the $n$-particle Green's function and yields $2$ (from the original two terms) times $2^n$ (number of possible combinations of terms from the binomials) terms of the structure
\begin{equation}
\big( \Gup - \Glow \big)^{"l"}  \, {G^{(1)}}^{"n - l"},
\label{struct}
\end{equation}
which have to be summed up. Note that the expression for a summand as given in Eq.~(\ref{struct}) represents a pure symbolic notation with the exponents ${}^{"l"}$ and ${}^{"n - l"}$ denoting the number of factors of a given type appearing in a single term, without explicitly specifying which frequencies ($\nu_1 , \nu_2, \nu_3, ...$) they are associated with. Obviously, each such term appears exactly twice in the sum [corresponding to the original two summands in Eq.~(\ref{Gn})], once with $p_1 (p_2)^l$ and once with $p_2 (-p_1)^l$ as a prefactor. Hence, we define the factor $\mathcal{F}_l$ (note that this $\mathcal{F}$ is not the full vertex, but a factor that will appear in the formula for $G^{(n)}$): 
\begin{equation}
\mathcal{F}_l = p_1 (p_2)^l + p_2 (-p_1)^l,
\label{Fl}
\end{equation}
which is associated with each summand consisting of a product of $l$ terms of type $(\Gup-\Glow)$ and $n-l$ terms of type $G^{(1)}$. This definition allows us to express $G^{(n)}$ in the following way:
\begin{multline}
\!\!G^{(n)} = \delta^{\nu_{1}...\nu_{n}}_{\nu_{n+1}...\nu_{2n}}\! \sum^{l = n}_{P , l = 0} \!  \mathcal{F}_l \, \big( \Gup - \Glow \big)(\nu_{P(1)}) \ldots \big( \Gup - \Glow \big)(\nu_{P(l)}) \\ \cdot G^{(1)}(\nu_{P(l+1)})G^{(1)}\ldots G^{(1)}(\nu_{P(n)}) \dfrac{1}{l! (n-l)!} ,\label{equ:gnrewrite}
\end{multline}
where $P$ denotes all permutations of the numbers $i=1\ldots n$. The factors $l!$ and $(n-l)!$ compensate for overcounting due to permutations which only exchange frequencies between the same type [$G^{(1)}$ or $( \Gup - \Glow )$] of term. For the sake of a better readability of the following derivations we introduce the following (symbolic) shorthand notation for the terms in the sum of Eq. (\ref{equ:gnrewrite}):
\begin{equation}
G^{(n)} = \sum^{n}_{l = 0}{\binom{n}{l}} \mathcal{F}_l \, \big( \Gup - \Glow \big)^{"l"}  \, {G^{(1)}}^{"n - l"} ,
\label{Gndecomp1}
\end{equation}
where the combinatorial factor ${\binom{n}{l}}$ indicates the summation over all ${\binom{n}{l}}$ permutations of the frequencies $\nu_1\ldots \nu_n$ in Eq. (\ref{equ:gnrewrite}).

\subsection{First part of the diagrammatic decomposition}
\label{sec:decomp1}

Eq.~(\ref{Gndecomp1}) represent an --up to this point algebraic-- decomposition of the full two-particle Green's function. Let us now analyze to which extent the latter already corresponds to a diagrammatic separation of terms. We first introduce the concept of one-particle disconnectedness: Diagrams in which a single Green's function line (with frequency $\nu$) is disconnected from the rest will be called one-particle disconnected (1PD) (in frequency $\nu$)\cite{note_1PD}. The form of Eq.~(\ref{Gndecomp1}) suggests that it already represents a decomposition of $G^{(n)}$ into 1PD diagrams and a remainder. In fact, the summands for $l=0\ldots n-1$ contain a product of $(n-l)$ one-particle Green's functions $G^{(1)}$ and a --not yet interpreted-- contribution $(\Gup-\Glow)^{"l"}$.  Hence, we assume that the terms for $l=1\ldots n-1$ contain {\sl all} 1PD diagrams while the remainder $l=n$ is one-particle connected (i.e., it exhibits no 1PD contributions). The latter will be denoted in the following as
\begin{figure}
\centering
\includegraphics[width=0.85\linewidth]{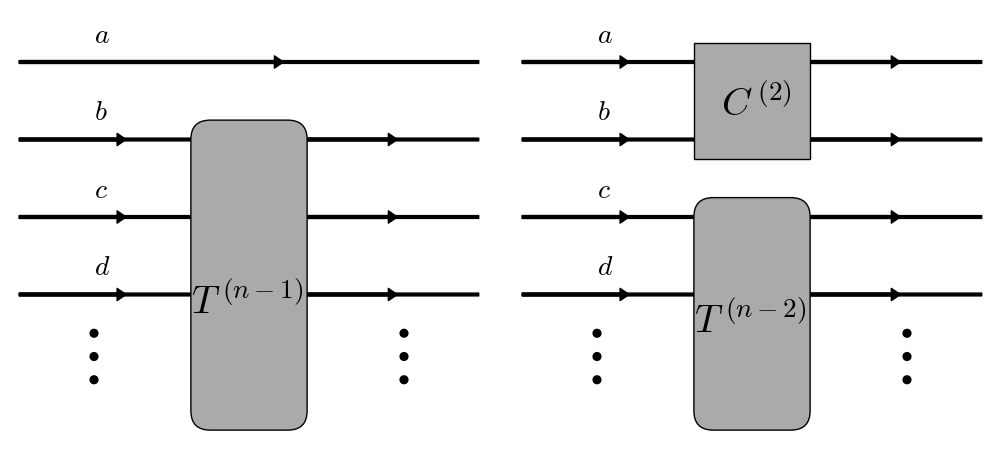}
\caption{(Color online) \label{exconn} Left: All diagrams where only the frequency $\nu_1$ is 1PD. Right: all the diagrams contributing to the $n$-particle one-particle-connected propagator $T^{(n)}$ where frequency $\nu_1$ is connected only to the frequency $\nu_2$.}
\label{fig:decomp}
\end{figure} 

\begin{multline}
T^{(n)}(\nu_1,\ldots,\nu_n,\nu_{n+1},\ldots \nu_{2n}) = \mathcal{F}_n \, \deltaABDE 
\\ \big( \Gup(\nu_1) - \Glow(\nu_1) \big) \ldots\big( \Gup(\nu_n) - \Glow(\nu_n) \big),
\label{claimT}
\end{multline}
and is illustrated diagrammatically in  Fig.~\ref{exconn} (left). We will now prove the above assumption, i.e., that $T^{(n)}$ indeed corresponds to the sum of all one-particle connected diagrams in $G^{(n)}$, by induction. Let us start from the base case $n=2$ for which in Ref. \onlinecite{Ribic16} we have calculated the full vertex. Since for $n=2$ all disconnected diagrams are 1PD, 
$T^{(2)}$ is the same as the vertex with externals legs (Green's function lines)  and reads [Eq.~(6) of Ref.~\onlinecite{Ribic16}]:
\begin{eqnarray}
 \nonumber &T^{(2)}& (\nu_1 , \nu_2 ;\nu_3 , \nu_4 ) \\ \nonumber &=& \delta^{\nu_1  \nu_2}_{\nu_3  \nu_4} \mathcal{F}_2 \big( \Gup (\nu_1) - \Glow (\nu_1) \big) \big( \Gup (\nu_2) - \Glow (\nu_2) \big)
\\ \nonumber &=& \delta^{\nu_1  \nu_2}_{\nu_3 \nu_4} p_1 p_2 \big( \Gup (\nu_1) - \Glow (\nu_1) \big) \big( \Gup (\nu_2) - \Glow (\nu_2) \big)
\\  &=& G^{(1)}G^{(1)}FG^{(1)}G^{(1)}
\label{T2}
\end{eqnarray}
with $F$ defined via the equation
\begin{eqnarray}
\nonumber G^{(2)}(\nu_1 , \nu_2 ;\nu_3 , \nu_4 ) = \delta^{\nu_1  \nu_2}_{\nu_3  \nu_4} G^{(1)} (\nu_1) G^{(1)} (\nu_2) 
\\+ \, G^{(1)}G^{(1)}FG^{(1)}G^{(1)}.
\end{eqnarray}
The frequency arguments of the expression $G^{(1)}G^{(1)}FG^{(1)}G^{(1)}$ have been omitted for brevity in both cases.
Obviously, Eq.~(\ref{T2}) demonstrates that our induction hypothesis, i.e., that $T^{(n)}$ corresponds to a sum of all one-particle connected diagrams, is correct for $n=2$. Moreover, from Eqs. (\ref{Gndecomp1}) and (\ref{claimT}) we have that
\begin{equation}
T^{(n+1)} =G^{(n+1)}-\sum^{n}_{l = 0} {\binom{n+1}{l}} T^{(l)}  \, {G^{(1)}}^{"n - l+1"}.
\label{Gndecomp1a}
\end{equation}
 Note that by the induction hypothesis, $T^{(l)}$ contains {\sl all} one-particle connected diagrams. Hence the term which is subtracted on the r.h.s. of Eq. (\ref{Gndecomp1a}) indeed is identical to the sum of {\sl all} 1PD diagrams in $G^{(n+1)}$. Consequently our induction step shows that $T^{(n+1)}$ does {\sl not} contain any 1PD diagram. Instead it includes all one-particle connected diagrams of $G^{(n+1)}$ which completes our inductive proof.  

Before we proceed with our decomposition procedure by removing the remaining disconnected contributions of $G^{(n)}$, let us point out some properties of the one-particle-connected $n$-particle propagator.
\begin{itemize}
\item For $n=2$ and $n=3$, $T^{(n)}$ already represents the fully connected contribution to $G^{(n)}$ as in these cases all disconnected diagrams are 1PD.
\item For odd $l$ and half-filling ($p_1 = p_2 = 1/2$), ${\cal F}_l=0$ and hence $T^{(l)} = 0 $ This also implies that the fully connected propagator and by extension the vertex have to vanish as will be discussed in the next section.
\end{itemize}

\subsection{Second part of the diagrammatic decomposition}
\label{sec:decomp2}

While the term $T^{(n)}$ derived in the previous section [see Eq. (\ref{claimT})] contains no 1PD diagrams, this does not automatically mean that it already corresponds to the fully connected part of the $n$-particle Green's function (as it is indeed the case for $n=2$ and $n=3$). For instance, $G^{(4)}$ contains --apart from 1PD parts-- a disconnected (but not 1PD) contribution proportional to $T^{(2)}T^{(2)}$ which is a product of one-particle connected diagrams and, hence, still included in $T^{(4)}$ (cf.\ Fig.~\ref{fig:decomp}).

In order to  obtain the fully connected part of $G^{(n)}$, which we will denote as $C^{(n)}$ in the following, we have to remove such contributions. This will be conducted in two steps: (i) We will derive the general functional form for $C^{(n)}$ leaving the appropriate prefactor still undefined. (ii) The prefactor will  then be determined in the second step by an explicit subtraction of all disconnected diagrams from $T^{(n)}$. 

(i) As for the first step we realize that for $n=2$ and $n=3$ the fully connected part of $G^{(n)}$ is proportional to $T^{(2)}$ and $T^{(3)}$, respectively, i.e., to $(\Gup-\Glow)^{"2"}$ and  $(\Gup-\Glow)^{"3"}$, as we have discussed above. This suggests the ansatz
\begin{align}
 \label{equ:ansatzconnect}
 C^{(n)}(\nu_1,&\ldots,\nu_n,\nu_{n+1},\ldots,\nu_{2n})=\nonumber\\&=\mathcal{C}_n\Bigl(\Gup(\nu_1)-\Glow(\nu_1)\Bigr)\ldots\Bigl(\Gup(\nu_{n})-\Glow(\nu_{n})\Bigr),
\end{align}  
where we have again suppressed the generalized Kronecker symbol $\deltaABDE$; $\mathcal{C}_n$ is a --for the moment-- unknown constant. We can now prove the assumption made in ansatz Eq.~(\ref{equ:ansatzconnect}) for the functional form of the fully connected $n$-particle Green's function by induction. 
The correctness for the base clause, i.e., for $n=2$, follows from Eq.~(\ref{T2}).  As induction hypothesis we consider that Eq.\ (\ref{equ:ansatzconnect}) holds for all $l=1\ldots n$, and  prove in the following that it then also holds for  $n+1$.

The fully connected $n+1$ particle Green's function $C^{(n+1)}$ can be obtained from the corresponding one-particle connected one $T^{(n+1)}$ (which we derived in the previous subsection) by removing all disconnected diagrams still present in this expression. Those disconnected diagrams can be written (due to the absence of 1PD contributions) as products of fully connected $1,2,\ldots,n$ particle Green's functions, i.e., $C^{(m_1)}\ldots C^{(m_k)}$ where $2<m_i\le n-1$ and $\sum_{i=1}^k m_i=n+1$. 
To obtain all terms that need to be  subtracted one has to sum over all possible Feynman diagrams of this type, i.e., over all partitions $m_i$ of $n+1$ and -for each such partition- over all possible frequency assignments to the single terms $C^{(m_i)}$ in the product. 

The crucial point is now, since by the induction hypothesis each of the $C^{(m_i)}\sim(\Gup-\Glow)^{``m_i``}$, that all the corresponding products in the sum are proportional to $\Bigl(\Gup(\nu_1)-\Glow(\nu_1)\Bigr)\ldots\Bigl(\Gup(\nu_{n+1})-\Glow(\nu_{n+1})\Bigr)$, where each of the incoming frequencies $\nu_1,\ldots,\nu_{n+1}$ is assigned to exactly one of the factors $(\Gup-\Glow)$. The only difference lies in the different prefactors of $\Bigl(\Gup(\nu_1)-\Glow(\nu_1)\Bigr)\ldots\Bigl(\Gup(\nu_{n+1})-\Glow(\nu_{n+1})\Bigr)$ for different terms in the sum. Moreover, since also $T^{(n+1)}$ (from which all these unconnected terms have to be subtracted) has the same structure [see Eq. (\ref{claimT})], the same must hold for $C^{(n+1)}$ which concludes our proof. 

(ii) Having identified the functional form of $C^{(n)}$ as given in Eq. (\ref{equ:ansatzconnect}) we are left with the task of determining the prefactor $\mathcal{C}_n$. To this end we will present a diagrammatic procedure that removes systematically  all disconnected contributions from the one-particle connected part of $G^{(n)}$, i.e., $T^{(n)}$. This will be achieved by choosing (without loss of generality) one ``reference''-frequency, in our case $\nu_1$, and classifying all subtracted (disconnected) diagrams with respect to this frequency.

Let us start with disconnected diagrams where the frequency $\nu_1$ is connected only to {\sl one} other frequency $\nu_2$ and the remaining frequencies $\nu_3,\nu_4,\ldots$ belong to all possible (one-particle-connected) ($n-2$)-particle diagrams. As was shown in Sec.~\ref{sec:decomp1} the latter corresponds exactly to $T^{(n-2)}(a_3,a_4,\ldots)$. The corresponding subtraction term is depicted in the right panel of Fig.~\ref{fig:decomp} and reads algebraically
\begin{align}
 \label{equ:decomp2ord}
\! C^{(2)}&(\nu_1,\nu_2)T^{(n-2)}(\nu_3,\nu_4,\ldots\!)\!=\!\mathcal{C}_2\mathcal{F}_{(n\!-\!2)}\Bigl(\!\Gup(\nu_1)\!-\!\Glow(\nu_1)\!\Bigr)\nonumber\\ \times &\Bigl(\!\Gup(\nu_2)\!-\!\Glow(\nu_2)\!\Bigr)\Bigl(\!\Gup(\nu_3)\!-\!\Glow(\nu_3)\!\Bigr)\Bigl(\!\Gup(\nu_4)\!-\!\Glow(\nu_4)\!\Bigr)\ldots,
\end{align}
where $\ldots$ denotes the multiplication with terms of the type $(\Gup-\Glow)$ for all remaining frequencies. In the same way a disconnected contribution, where the frequency $\nu_1$ is connected only to frequency $\nu_3$ (but disconnected from all the others), has to be removed from $T^{(n)}$ in order to obtain $C^{(n)}$, i.e., the term $C^{(2)}(\nu_1,\nu_3)T^{(n-2)}(\nu_2,\nu_4,\ldots)$. The corresponding explicit expression is, however, equivalent to the one on the r.h.s. of Eq.~(\ref{equ:decomp2ord}). The same obviously holds when applying the above procedure to all remaining frequencies $\nu_4,\ldots$ and, hence, the total subtraction term originating from diagrams where the frequency $\nu_1$ is connected to {\sl only one} other frequency is given by $n-1$ (i.e., the number of frequencies to which $\nu_1$ can be connected) times the expression on the r.h.s. in Eq.~(\ref{equ:decomp2ord}).

In the next step we will remove all diagrams from $T^{(n)}$ where the frequency $\nu_1$ is connected to {\sl two} other frequencies but disconnected to the rest such as, e.g., $C^{(3)}(\nu_1,\nu_2,\nu_3)T^{(n-3)}(\nu_4,\nu_5,\ldots)$. Following the arguments from the previous paragraph, this yields $(n-1)(n-2)/2$ (corresponding to the number of ways to select $2$ frequencies from a set of $n-1$ frequencies) equivalent terms of the form $\mathcal{C}_3\mathcal{F}_{(n-3)}(\Gup(\nu_1)-\Glow(\nu_1))\ldots$ which have to be removed from $T^{(n)}$. 

Extending the above arguments to one-particle disconnected diagrams where a number $1 \le l-1 \le n-3$ frequencies are connected to the frequency $\nu_1$ and all the others are disconnected from $\nu_1$, we obtain a subtraction term of the form ${\binom{n-1}{l-1}}\mathcal{C}_l\mathcal{F}_{(n-l)}(\Gup(\nu_1)-\Glow(\nu_1))\ldots$. We can, hence, obtain an explicit expression for $C^{(n)}(\nu_1,\nu_2,\nu_3,\ldots)$ by subtracting all the above mentioned contributions from the one-particle connected function $T^{(n)}(\nu_1,\nu_2,\nu_3,\ldots)$ which yields
\begin{align}
\mathcal{C}_n&\Bigl(\Gup(\nu_1)-\Glow(\nu_1)\Bigr)\ldots  = \mathcal{F}_{n}\Bigl(\Gup(\nu_1)-\Glow(\nu_1)\Bigr)\ldots\nonumber\\& - \sum^{n - 2}_{l = 2} {\binom{n-1}{l-1}} \mathcal{C}_l \mathcal{F}_{n-l} \Bigl(\Gup(\nu_1)-\Glow(\nu_1)\Bigr)\ldots \;\;.
\label{Cequationiter2}
\end{align}
Since the product of $(\Gup-\Glow)$ appears for each summand in this equation it can be removed in each term. This finally yields the following iterative expression for the prefactor $\mathcal{C}_n$
\begin{equation}
\mathcal{C}_n = \mathcal{F}_n - \sum^{n - 2}_{l = 2} {\binom{n-1}{l-1}} \mathcal{C}_l \mathcal{F}_{(n-l)}
\label{Cequationiter}
\end{equation}
After the determination of $\mathcal{C}_n$ from the above relations, the connected $n$-particle Green's function is given by Eq.~(\ref{equ:ansatzconnect}).

In a final step we can now calculate the connected $n$-particle vertex function $F^{(n)}$ from $C^{(n)}$ by just amputating the external legs, i.e., via division by the corresponding one-particle Green's functions $G^{(1)}$. Defining $f(\nu)$
\begin{equation}
f(\nu) = \big( \Gup (\nu) - \Glow (\nu) \big) / \big( G^{(1)} (\nu) \big)^2,
\label{EPA}
\end{equation}
the explicit expression for the full $n$-particle vertex $F^{(n)}$ is given by
\begin{multline}
F^{(n)} (\nu_1,\ldots,\nu_n,\nu_{n+1},\ldots,\nu_{2n}) \\ = \deltaABDE \mathcal{C}_n  f(\nu_1) \ldots f(\nu_n).
\label{nVert}
\end{multline}
For $n=2$ this is equivalent to the corresponding expression found in Ref.~\onlinecite{Ribic16}.

\section{Investigating the properties of $\mathcal{C}_n$}
\label{sec:propC}

\begin{figure}
\centering
\includegraphics[width=0.85\linewidth]{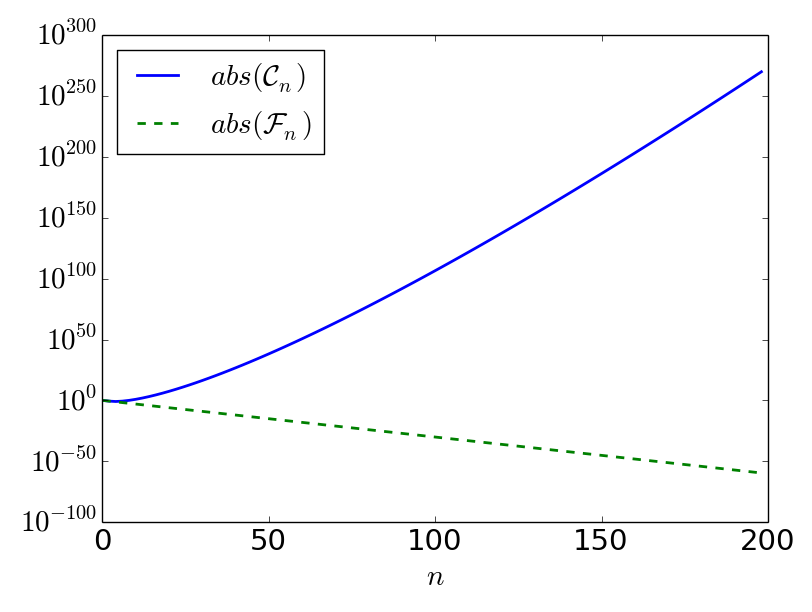}
\caption{(Color online) Coefficients $abs(\mathcal{C}_n)$ ($abs(\mathcal{F}_n)$) for the (one-particle) connected $n$-particle vertex at half-filling $p_1=p_2 = 0.5$ for even $n$ up to $200$. Note the logarithmic scale.
\label{upto200halfln}}
\end{figure}

\begin{figure}
\centering
\includegraphics[width=0.85\linewidth]{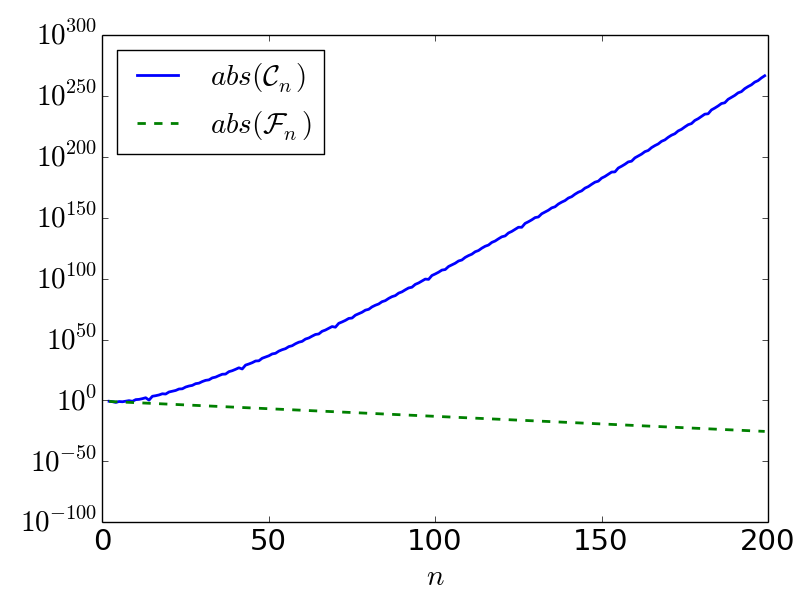}
\caption{(Color online) Same as Fig. \ref{upto200halfln}, but for quarter-filling $p_1 = 0.25$ and also including odd numbers for $n$.
\label{upto200quarterln} }
\end{figure}

In this section we want to analyze the behavior of the prefactor $\mathcal{C}_n$ for increasing $n$ which can be considered as a first estimate for the ``magnitude'' of the vertex function of a certain order $n$. This is a relevant issue for various diagrammatic extensions of DMFT such as DF and 1PI which typically construct nonlocal corrections to the DMFT self-energy from the two-particle local vertex function assuming that all higher-order vertices are in some sense ``small''. Hence, in order to obtain an understanding of the magnitude of $\mathcal{C}_n$ we iterate Eq.~(\ref{Cequationiter2}) up to a given order $n>2$ starting from the initial conditions $\mathcal{F}_2=\mathcal{C}_2=p_1p_2$ (recall that $p_1$ denotes the density of localized $f$ electrons and $p_2=1-p_1$). 
\begin{figure}
\centering
\includegraphics[width=0.85\linewidth]{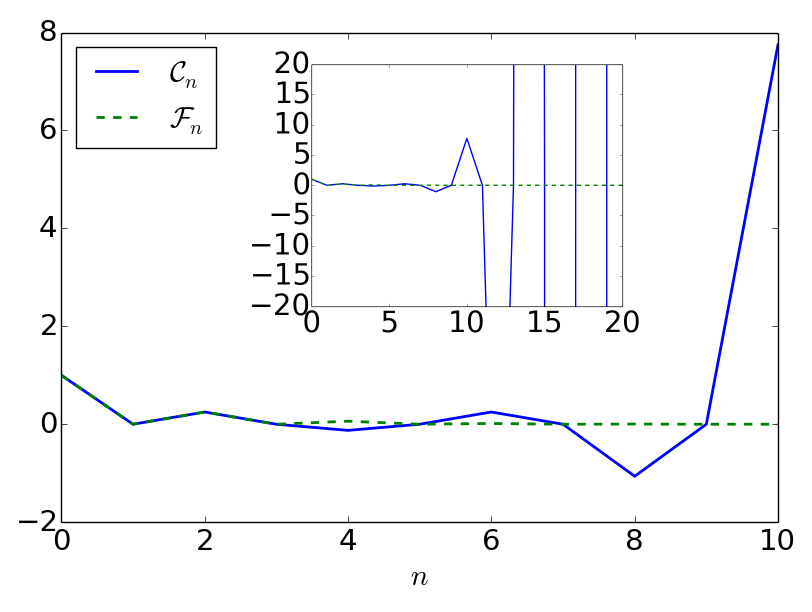}
\caption{(Color online)
\label{upto10half} Coefficients $\mathcal{C}_n$ ($\mathcal{F}_n$) for the connected (one particle connected) $n$-particle Green's function at half-filling $p_1=p_2 = 0.5$ for $n$ up to $10$. The insert shows the behavior up to $n = 20$.}
\end{figure} 
\begin{figure}
\centering
\includegraphics[width=0.85\linewidth]{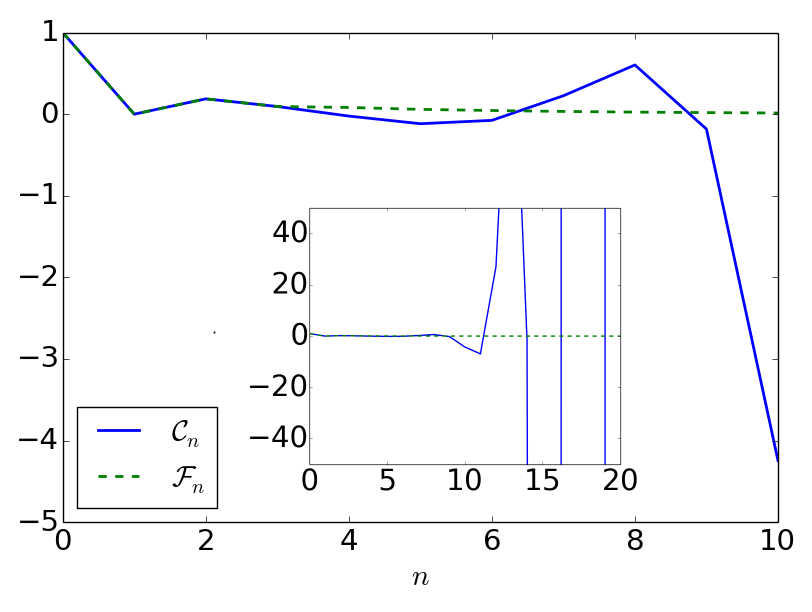}
\caption{(Color online) 
\label{upto10quarter} Same as Fig.\ \ref{upto10half} but for quarter filling  $p_1 = 0.25$.}
\end{figure} 
In Figs. \ref{upto10half} and \ref{upto10quarter} we present our data for $\mathcal{C}_n$ and $\mathcal{F}_n$ up to $n=10$ for half\cite{halffilling} ($p_1=p_2=0.5$) and quarter ($p_1=0.25,p_2=0.75$) filling, respectively. For these small values of $n$, the size of $\mathcal{C}_n$ is still moderate. 
However, when extending the range $n$ up to $n=20$ (for half-filling) we observe already a much stronger increase of $\mathcal{C}_n$ with $n$ (see inset of Fig. \ref{upto10half}). By increasing $n$ further up to $n = 200$ (see Fig. \ref{upto200halfln} for half- and Fig. \ref{upto200quarterln} for quarter-filling) we find an extremely strong increase of $\mathcal{C}_n$: In fact, plotting  $\log \lvert \mathcal{C}_n\rvert$ (rather than $\mathcal{C}_n$ itself) we observe an increase even stronger than linear on this logarithmic scale in  Figs.~\ref{upto200halfln} and \ref{upto200quarterln}. This indicates that $\mathcal{C}_n$ is  growing faster than exponentially with $n$.

\begin{figure}
\centering
\includegraphics[width=0.85\linewidth]{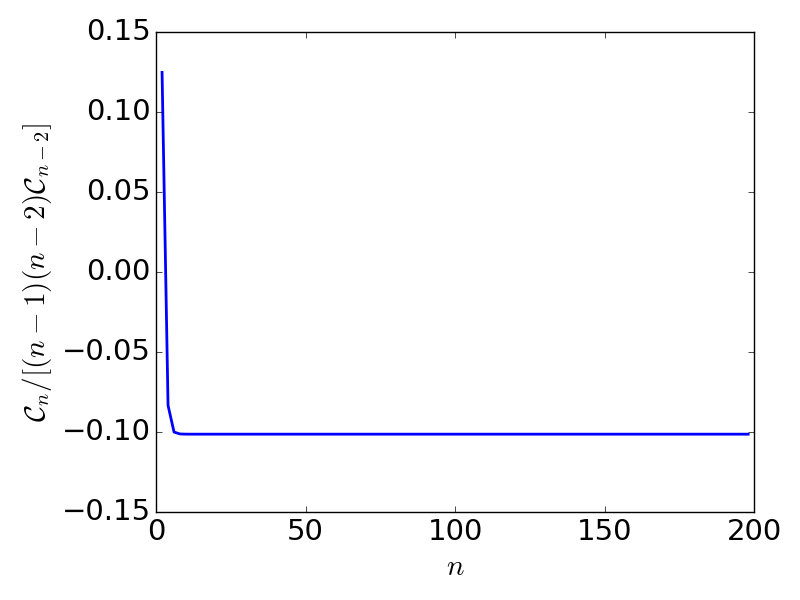}
\caption{(Color online) Ratio $r_n=\mathcal{C}_n/[(n-1)(n-2)\mathcal{C}_{n-2}]$ at half-filling $p_1=p_2 = 0.5$. Note that $r_n$ converges towards a negative constant, implying an alternating sign of $\mathcal{C}_n$ (for even $n$). }
\label{halfratio}
\end{figure}

Let us provide an analytical understanding for this numerically observed growth rate of $\mathcal{C}_n$ in the simple case of half-filling ($p_1=p_2=1/2$) where $\mathcal{C}_n\equiv 0$ for all odd\cite{halffilling} $n$ and, hence, the sum in Eq.~(\ref{Cequationiter2}) is restricted to even $l$. The binomial coefficient in this relation suggests a factorial growth of $\mathcal{C}_n$: In fact, considering the last term in this sum we obtain that $\mathcal{C}_n\propto (n-1)(n-2)\mathcal{C}_{n-2}$. Hence, we expect that, $\mathcal{C}_n/[(n-1)(n-2)\mathcal{C}_{n-2}]$ of two neighboring (non-zero) coefficients in the recursion relation of Eq.~(\ref{Cequationiter2}) should be constant for large values of $n$, i.e., 
\begin{equation}
\label{equ:defr}
r_n = \frac{\mathcal{C}_n}{\mathcal{C}_{n-2}(n-1) (n-2)},
\end{equation}
for $n\rightarrow \infty$. Fig. \ref{halfratio} shows the behavior of $r_n$ as a function of $n$. We can clearly see that it approaches a constant value which can be numerically estimated as $r_{n\rightarrow\infty}=r_a \approx -0.1$. Note that the minus sign of this number reflects perfectly the alternating behavior of the sign of $\mathcal{C}_n$ for neighboring (even) $n$ observed in Fig.~\ref{upto10half}.

Eq.~(\ref{equ:defr}) can be now used as an iteration scheme which allows us to find an (approximate) explicit expression for $\mathcal{C}_n$ at large values of $n$. It is given by
\begin{equation}
\label{equ:capprox}
\mathcal{C}_n \underset{n \rightarrow \infty}{=} K \sqrt{r_a }^{n-2} \, (n-1)!,
\end{equation}
with some prefactor $K$ (which should be fitted to the large-$n$ tail of $\mathcal{C}_n$). We next insert the above form of $\mathcal{C}_n$ into both sides of the recursion relation in Eq.~(\ref{Cequationiter2}) which yields (considering the simplifications occuring for $n\rightarrow\infty$) $r_a=-1/\pi^2\sim-0.1$, coinciding exactly with the numerical prediction.

\begin{figure}
\centering
\includegraphics[width=0.85\linewidth]{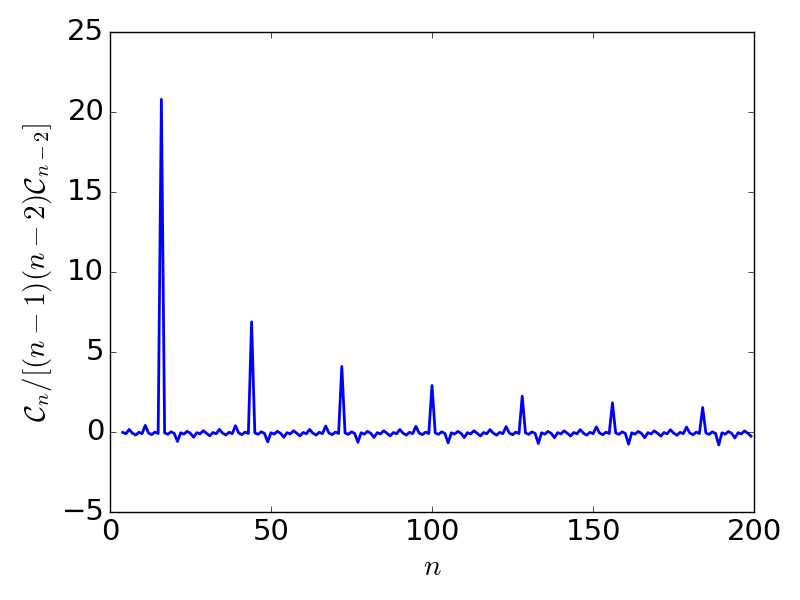}
\caption{(Color online)\label{quarterratio} Same as Fig. \ref{halfratio}, but for $p_1 = 0.25$. Away from half-filling, quasi-periodic structures arise in the ratio.}
\end{figure}

Eq. (\ref{equ:capprox}) demonstrates that $\mathcal{C}_n$ grows --at half-filling-- factorially with $n$. Let us, however, point out here that the above considerations and approximations cannot be directly transferred to situations out of half-filling due to the emergence of quasi-periodic structures in $r$ as it can be observed in Fig.~\ref{quarterratio}.

At a first glance the strong increase of the magnitude of the local DMFT vertex functions with the particle number might indeed invalidate the state-of-the-art diagrammatic extensions of DMFT such as DF and 1PI in their standard formulation, where they are restricted to the two-particle vertex level. Two further aspects, however, have to be considered which put the above argument in perspective: (i) The alternating behavior of the sign of $\mathcal{C}_n$ (see Figs. \ref{upto10half} and \ref{upto10quarter}) will certainly mitigate a strong growth of diagrammatic contributions upon increasing $n$ in the aforementioned methods, and (ii) the larger number of Green's function in diagrams with higher order vertices might compensate the the increasing size of the vertex functions itself. The later issue will be discussed in more detail in the following section. 


\section{Estimate of higher order vertex contributions}
\label{sec:ordermagnitude}
\begin{figure}
\centering
\includegraphics[width=0.95\linewidth]{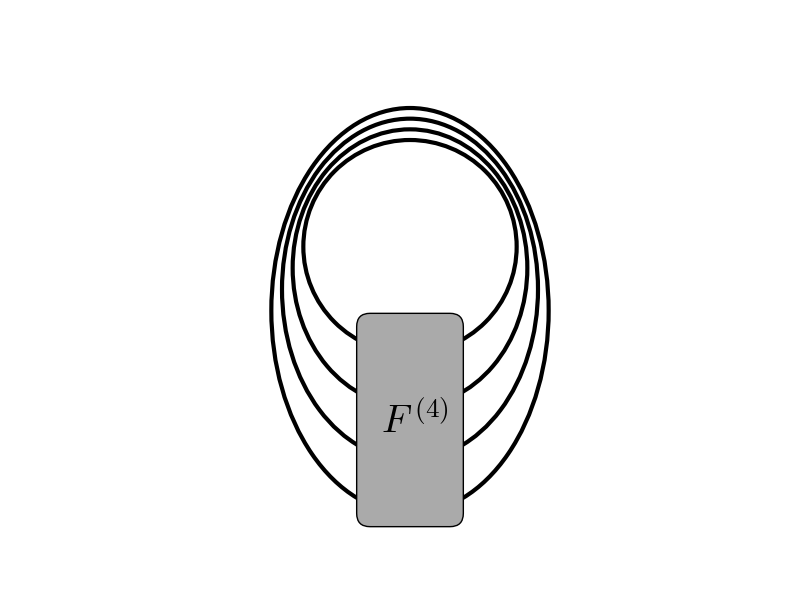}
\caption{(Color online)\label{Konvvert} An exemplary diagram (``necklace'' diagram) contributing to the partition function, which can be generated within a diagrammatic extension of DMFT such as DF or 1PI including the local 4-particle vertex function. Generalizations to corresponding diagrams with higher-order local vertices are straightforward.}
\end{figure}

In the following we will estimate how the factorial growth of the local $n$-particle vertex of the FKM discussed in the previous section will affect the magnitude of Feynman diagrams for the partition function, which are constructed from such vertices in the framework of diagrammatic extensions of DMFT. To this end we will evaluate a generic partition function diagram assuming a general atomic-limit-like form of the Green's functions within the dual fermion framework. Within dual fermion theory, the action of the full lattice system is systematically decoupled into local and non-local degrees of freedom. The starting point is the action of the Falicov-Kimball model, expressed in the (imaginary) time-dependent Grassmann fields $c^{(+)}$ and $f^{(+)}$ that correspond to the $c^{(\dag)}$ and $f^{(\dag)}$ of the Hamiltonian (\ref{equ:defhamilt}):
\begin{align}
\mathcal{S}[c^+,c,f^+ ,f]=&\sum_{\mathbf{k}\nu}\left[-i\nu+\varepsilon_{\mathbf{k}}-\mu_c\right]c^+_{\mathbf{k}\nu}c^{\phantom +}_{\mathbf{k}\nu}\nonumber \\ + &\sum_{i\nu} \left[-i\nu-\mu_f\right]f^+_{i\nu}f^{\phantom +}_{i\nu}\nonumber \\ +U&\sum_i\int_0^{\beta}d\tau\;c^+_{i}(\tau)c^{\phantom \!}_{i}(\tau)f^+_{i}(\tau)f^{\phantom \!}_{i}(\tau),
\end{align}
Here, $\varepsilon_{\mathbf{k}}$ gives the dispersion relation of the non-interacting system. A hybridisation function $\Delta_{\nu}$ is added to and subtracted from the action. 
\begin{align}
\label{locnonloc}
\mathcal{S}[c^+,c,f^+ ,f]=\sum_{i}\mathcal{S}_{\text{loc}}[c_{i}^+,c^{\phantom \!}_{i},f^+_i ,f^{\phantom \!}_i]\nonumber\\+\sum_{\mathbf{k}\nu}\left[\varepsilon_{\mathbf{k}}-\Delta_{\nu}\right]c^+_{\mathbf{k}\nu}c^{\phantom +}_{\mathbf{k}\nu}.
\end{align}
This hybridization is a quadratic term in the fermionic $c^{(+)}$ fields and allows us to identify the action of our referential DMFT problem.
\begin{align}
\mathcal{S}_{loc}[c^+,c,f^+ ,f]=&\sum_{\nu}\left[-i\nu+\Delta_{\nu}-\mu_c\right]c^+_{\nu}c^{\phantom +}_{\nu}\nonumber \\ + &\sum_{\nu} \left[-i\nu-\mu_f\right]f^+_{\nu}f^{\phantom +}_{\nu}\nonumber \\ +U&\int_0^{\beta}d\tau\;c^+(\tau)c(\tau)f^+(\tau)f(\tau).
\end{align}
The difference between kinetic energy and hybridisation function $(\varepsilon_{\mathbf{k}}-\Delta_{\nu})$ in Eq.~(\ref{locnonloc}) couples the different local DMFT problems. In dual fermion, we next employ a Hubbard-Stratanovich transformation. This way the problem is recast into a system of new, fermionic, degrees of freedom, the dual fermions $\widetilde{c}^{(+)}$. The local problems are then integrated by solving the associated DMFT problems. Due to the Hubbard-Stratanovich transformation, the dual fermions couple to the original $c$ ones. Therefore, upon integrating out the local problems, one ends up with an infinite series of DMFT $n$-particle Green's functions that are coupled to the dual fields (up to a multiplicative factor). In writing the solution to the local problems as an exponential function, the DMFT vertices formally take the role of a bare interaction for the dual fermions. Mathematically, we obtain after the Hubbard-Stratanovich transformation the following action (For a more detailed derivation see Ref. \onlinecite{DualFermion,RohringerPhD,RMP}):
\begin{align}
\label{dualaction}
\mathcal{S}[\widetilde{c}^+,\widetilde{c}] = \sum_{i} \sum_{n=2}^\infty (-1)^n F^{(n)} [\widetilde{c}_{i}^+,\widetilde{c}_{i}] \nonumber+\nonumber \\ \sum_{\mathbf{k}\nu}\left( G_\mathbf{k} (\nu) - G_{loc} (\nu) \right)^{-1}\widetilde{c}^+_{\mathbf{k}\nu}\widetilde{c}^{\phantom +}_{\mathbf{k}\nu}.
\end{align}
Here, each $n$-particle vertex $F^{(n)}$ is multiplied by the dual Grassmann fields associated with the site $i$ and the Matsubara frequencies of the entering and leaving fermions of $F^{(n)}$. $G_\mathbf{k} (\nu) = \left( i \nu - \varepsilon_{\mathbf{k}} - \Sigma (\nu) + \mu_c\right)^{-1}$ is the DMFT lattice Green's function. The action (\ref{dualaction}) can now be evaluated in terms of Feynman diagrams that consist of $n$-particle interactions $F^{(n)}$ (Hitherto truncated after $F^{(2)}$) and non-interacting Green's functions 
\begin{equation}
\GDF_0 (\mathbf{k} , \nu) = \langle \widetilde{c}^+_{\mathbf{k}\nu}\widetilde{c}^{\phantom +}_{\mathbf{k}\nu} \rangle_0 = G_\mathbf{k} (\nu) - G_{loc} (\nu).
\end{equation}
After this synopsis of DF, let us turn back to estimating  typical Feynman diagrammatic contributions.
To  estimate the convergence behavior of generic diagrams,  we employ an asymptotic (high frequency) expression for the dual fermion Green's function
\begin{equation}
\GDF_{\text{approx}}(\nu_l) \propto \dfrac{\overline{\varepsilon}}{\nu_l^2},
\end{equation}
where $\nu_l= \pi T (2l - 1)$, $l\in\mathds{Z}$. Note that the dual Fermion Green's function is the difference of the ${\mathbf k}$-dependent and a local Green's function $G_\mathbf{k} (\nu) - G_{loc} (\nu)$ in action (\ref{dualaction}) so that the asymptotic high frequency behavior is $\sim1/\nu_l^2$;  $\overline{\varepsilon}$ is a characteristic energy scale of the system. The simplest diagrams for the partition function which can be constructed from this Green's function and the $n$-particle vertex functions of DMFT are of first order in the $n$-particle vertex: They can be obtained by connecting all its entrances and exits pairwise by $n$ one-particle Green's functions $\GDF_{\text{approx}}$. An example of such a diagram adopting the $n=4$-particle vertex is depicted in Fig.~\ref{Konvvert}. In the following we will analyze the contribution of such diagrams to the partition function, which we will denote as $\Omega^{(n)}$, for large values of $n$. Note that while the bare dual propagator does not have a local component, in a self consistent dual fermion formulation, local contributions can appear due to self-energy corrections. The basic structure of these terms is given by\cite{note_prefactor_diagram}
\begin{align}
 \label{equ:gendiagram}
 \Omega^{(n)} \sim \sum_{\nu_1\ldots\nu_{2n}}&\GDF_{\text{approx}}(\nu_1)\ldots \GDF_{\text{approx}}(\nu_n)\nonumber\\\times& F^{(n)}(\nu_1,\ldots,\nu_{2n}).
\end{align}
For this sum we have to keep in mind that the vertex $F^{(n)}$ vanishes if two (incoming) frequency arguments $a_1\ldots a_n$ take the same value. Consequently the largest contribution in the frequency sum in Eq.~(\ref{equ:gendiagram}) is the one where each of the lowest $n/2$ positive and negative  Matsubara frequencies $\pm \nu_1\ldots \pm\nu_{n/2}$ is attached to exactly one of the $n$ Green's functions. Assuming that $n$ is even we can, hence, estimate the dominant contribution of the diagram in Fig.~\ref{Konvvert} as
\begin{equation}
\Omega^{(n)}\sim U^n \prod_{l = 1}^{n/2} \left( \dfrac{\overline{\varepsilon}}{\nu_l^2} \, \dfrac{\overline{\varepsilon}}{\nu_l^2} \right) \mathcal{C}_n = \dfrac{\beta^{2n} \overline{\varepsilon}^n U^n}{(2 \pi )^{2n}} \prod_{l = 1}^{n/2} \dfrac{1}{(2l - 1)^4} \mathcal{C}_n,
\label{Gapprox}
\end{equation}
where we have replaced the full vertex $F^{(n)}$ just by its prefactor $\mathcal{C}_n$ times $U^n$, which originates from the limiting value of $f(a)$ for high frequencies. This is justified for a metallic like Green's function since the contribution $1/[G^{(1)}(\nu)]^2$, present in the definition of $F^{(n)}$ in Eq.~(\ref{nVert}), is non-zero and remains bounded for all frequencies.

Inserting now the asymptotic form of $\mathcal{C}_n$ as given in Eq.~(\ref{equ:capprox}) into Eq.~(\ref{Gapprox}) yields (after some simple algebraic manipulations)
\begin{equation}
\Omega^{(n)} \sim K (U \beta^2 \overline{\varepsilon})^n \dfrac{ \sqrt{ r_a } ^{n-2}  }{ \pi ^{2n}  }  \dfrac{\left[    (n/2 - 1)! \right]^4}{\left( (n - 1)!\right)^3}
\end{equation}
The application of Stirling's formula allows us to get rid of the factorial expressions.
\begin{equation}
\Omega^{(n)} \propto 4 K (U \beta^2 \overline{\varepsilon})^n \dfrac{ \sqrt{ r_a }^{n-2}  }{ (2\pi )^{2n}  }  \dfrac{1}{n^{n-1} e^n} .
\label{equ:omegapprox}
\end{equation}
This shows that even though the prefactor of ${\cal C}_{n}$ is growing postexponentially, the corresponding contribution to the partition function gets damped and is decaying.
The  sum over $n$ of the terms Eq.~(\ref{equ:omegapprox})  is even absolutely convergent.

Note however that our approximation is only valid for the case $T \gtrsim D$, with $D$ being the bandwidth of the system, as it relies on the asymptotic values of the vertex and Green's function being reached. Certainly, the limit $T \rightarrow 0$ cannot be taken. On one hand, this would violate our assumption that the first summand in terms of frequencies gives the dominant contribution to the sum, on the other hand the specific approximation employed for the Green's functions is bound to introduce divergencies in such a case.


Finally, a remark is in order about other types of diagrams for the partition function than the ones discussed above. To this end let us consider, e.g., a diagram where a second $n$-particle vertex is inserted. This leads to a factor $\mathcal{C}_n^2$ rather than $\mathcal{C}_n$. However, also the number of the Green's functions in the diagram is doubled and, hence, the corresponding diagram behaves like the square of our evaluated value. In general, we can state that the diagram depicted in Fig.~\ref{Konvvert} and corresponding higher-order diagrams represent somehow the ``worst case'' regarding the convergence with large $n$. 
Hence, although the above calculation has been done for the rather specific case of the FKM using the vertex of the half-filled system, it might be seen as a justification of the general restriction of the DF expansion to the lowest order vertex functions. This, however, does not guarantee that contributions for low $n>2$ are fully negligible, as we will demonstrate in the following section by including a $n=3$-particle diagram for the calculation of DF self-energy corrections in the FK model.

\FloatBarrier

\section{Numerical results for $3^{rd}$ order terms within dual fermion theory}
\label{sec:numerical}

\begin{figure}
\centering
\includegraphics[width=0.75\linewidth]{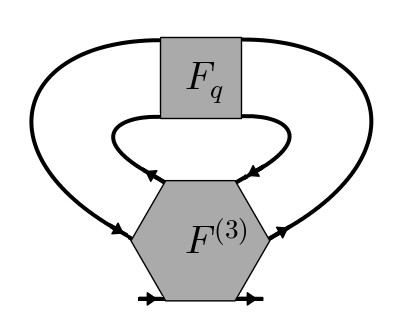}
\caption{\label{3rdCorrDiag}Diagrammatic representation of correction to the dual $\Sigma$ due to a $3$-particle vertex.}
\end{figure}

In this section we present numerical results for the non-local corrections to the DMFT self-energy constructed from two- and three-particle local DMFT vertices within the DF approach. To this end we have considered the FKM at quarter $f$-filling ($p_1$ = $0.25$) for the parameters $U = 1$, $\mu_c = 0.2$ and a temperature of $0.05$. For these parameters, we have a $c$-electron filling of $\langle c^\dag c \rangle \approx 0.53$ within DMFT. We now explicitly calculate  the  diagrams with a three-particle vertex  that is depicted in Fig.~\ref{3rdCorrDiag}. This contribution is beyond DMFT and beyond standard DF which is truncated at the two-particle vertex level.  In the diagram Fig.~\ref{3rdCorrDiag}, four of the six outer legs of the local DMFT three-particle vertex $F^{(3)}$ of the FKM are connected with a ladder built from two-particle vertices as depicted in Fig.~\ref{3rdCorrDiag}. 
Since $F^{(3)}$ is purely local the corresponding self-energy correction will be also $\mathbf{k}$-independent, similar to the local correction term appearing within a 1PI calculation. The latter, in fact, represents just the contribution of Fig.~\ref{3rdCorrDiag} where instead of the full $F^{(3)}$  its one-particle reducible part is considered [see Ref.~\onlinecite{1PI}].

\begin{figure}
\centering
\includegraphics[width=0.55\linewidth]{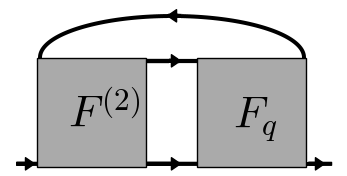}
\caption{\label{2ndCorrDiag}Diagrammatic representation of the (local) dual $\Sigma$ on the two-particle level.}
\end{figure}

Unlike the 1PI-case however, the correction  Fig.~\ref{3rdCorrDiag} to the local self-energy applies to the dual fermions. This dual self-energy has to be mapped to the real fermion self energy\cite{DualFermion}. This mapping  intermixes different contributions to the dual self-energy and
 affects the self-energy of the real fermions for different ${\mathbf k}$-points in slightly different ways. In the following we will however concentrate on the  self-energy in the dual space only, since only in dual space we can clearly distinguish contributions from the three-particle vertex and the standard ladder diagrams.
The standard dual fermion self-energy\cite{DualFermion,RohringerPhD,RMP} as shown in Fig.~\ref{2ndCorrDiag} is given by
\begin{eqnarray}
\label{equ:2ordcorr}
\Sigma_{\rm DF}( k ) = - \sum_{k_1,q} F^{(2)\; \nu \nu_1 \omega} \GDF (k_1) \GDF (k_1\!+\!q) \nonumber \\ \times F_{DF}^{k_1, k, q} \GDF ( k\!+\!q) . 
\end{eqnarray}
Here, $\GDF(k)=G^{\rm DMFT}_k-G_{\rm loc}(\nu)$ ($G_{\rm loc}(\nu)=\sum_{\mathbf k} G^{\rm DMFT}_k$) is the dual fermion Green's function which is given by the non-local part of the DMFT lattice Green's function $G^{\rm DMFT}_k$ \cite{DualFermion,Ribic16}, and $k=(\nu,{\mathbf k})$ is a compound momentum and frequency index. $F^{(2)}$ is the local two-particle vertex 
which only depends on the frequency components, written in $ph$-notation; $F_{DF}$ 
is the full momentum and frequency dependent vertex of the dual Fermions, approximated diagrammatically as a ladder built from  $F^{(2)}$, see below.

The explicit form of the dual self-energy corrections generated by the diagram with a three-particle vertex $F^{(3)}$  in Fig.~\ref{3rdCorrDiag} is, on the other hand, given by
\begin{eqnarray}
\label{equ:3ordcorr}
\Sigma^3_{\rm DF}( k ) = - \dfrac{1}{4} \sum_{k_1,k_2,q} F^{(3)} ( \nu\!, \nu_1\!, \nu_2\!+\!\omega\!, \nu\!, \nu_1\!+\!\omega\!, \nu_2 ) \nonumber \\ \times \GDF ( k_1 ) \GDF ( k_1\!+\!q ) F_{DF}^{k_1, k_2, q} \GDF ( k_2 ) \GDF ( k_2\!+\!q ) .
\end{eqnarray}
 For $F^{(3)}$ we will adopt the expression derived previously in Eq.~(\ref{nVert}) for $n=3$. The factor in front of the sum in Eq.~(\ref{equ:3ordcorr}) accounts for the overcounting due to the exchangeability of the lines entering and leaving the two-particle vertex.

Let us now consider the definition of the generalized susceptibility
\begin{equation}
\overset{\sim}{\chi}_{0,q}^{\nu_1 , \nu_2} = \sum_{k_1} \GDF ( \nu_1 , k_1 ) \GDF ( \nu_2 , k_1 + q ) ,
\end{equation}
and the particle-hole ladder with the local $F^{(2)}$ as a building block which yields
\begin{eqnarray}
 F_q^{\nu , \nu' , \omega} = F^{(2) \; \nu , \nu' , \omega} - \sum_{\nu_1} F^{(2) \; \nu , \nu_1 , \omega} \, \overset{\sim}{\chi}_{0,q}^{\nu_1 , \nu_1+\omega} F_q^{\nu_1 , \nu' , \omega} .
\end{eqnarray}
Considering that the contributions from the diagrams built employing the particle-hole ladder and the transversal particle-hole ladder to the dual self-energy are equal, we can replace $F^{\rm DF}\rightarrow(2 F_q - F^{(2)})$, where a double counted term had to be compensated for \cite{Ribic16}.

Altogether this gives  the  following explicit expression for Eq.~(\ref{equ:3ordcorr})
\begin{eqnarray}
\Sigma^3( \nu ) &&= - \dfrac{1}{2} \sum_{\nu_1 , \nu_2 , \omega , q} F^{(3)} ( \nu , \nu_1 , \nu_2 + \omega , \nu , \nu_1 + \omega , \nu_2 ) \nonumber \\ && \times  \overset{\sim}{\chi}_{0,q}^{\nu_1 , \nu_1 + \omega}  \left(\! F_q^{\nu_1 , \nu_2 , \omega} - \dfrac{1}{2} F^{(2) \; \nu_1 , \nu_2 , \omega} \!\right)   \overset{\sim}{\chi}_{0,q}^{\nu_2 , \nu_2 + \omega},
\end{eqnarray}
 Utilizing the frequency structure, especially the factorization of the local three-particle vertex $F^{(3)}$, we can now express the self-energy correction in terms of two auxiliary matrices
\begin{eqnarray}
X(\nu_1 , \nu_2) &=& f(\nu_1) f(\nu_2) \sum_q \overset{\sim}{\chi}_{0,q}^{\nu_1 , \nu_1} \, \left( F_q^{\nu_1 , \nu_2 , 0} - \dfrac{1}{2} F_{loc}^{\nu_1 , \nu_2 , 0} \right) \nonumber \\ && \times  \, \overset{\sim}{\chi}_{0,q}^{\nu_2 , \nu_2}
\end{eqnarray}
\begin{eqnarray}
Y(\nu_1 , \nu_2) &=& - f(\nu_1) f(\nu_2) \sum_q \overset{\sim}{\chi}_{0,q}^{\nu_1 , \nu_2} \, \Big( F_q^{\nu_1 , \nu_1 , \nu_2 - \nu_1} \nonumber \\ &&   - \dfrac{1}{2} F_{loc}^{\nu_1 , \nu_2 , \nu_2 - \nu_1} \Big) \, \overset{\sim}{\chi}_{0,q}^{\nu_1 , \nu_2}
\end{eqnarray}
For calculating the self-energy corrections, all admissible values of $\nu_1$ and $\nu_2$ have to be summed over
\begin{equation}
\Sigma^3( \nu ) = \mathcal{C}_3 f(\nu) \!\!\!\!\!\!\sum_{\nu_1 , \nu_2 | \nu \neq \nu_1 \neq \nu_2 \neq \nu}\!\!\!\!\!\! [ X(\nu_1 , \nu_2) + Y(\nu_1 , \nu_2) ].
\end{equation}
\begin{figure*}
\begin{minipage}{.9\textwidth}
\includegraphics[width=0.95\linewidth]{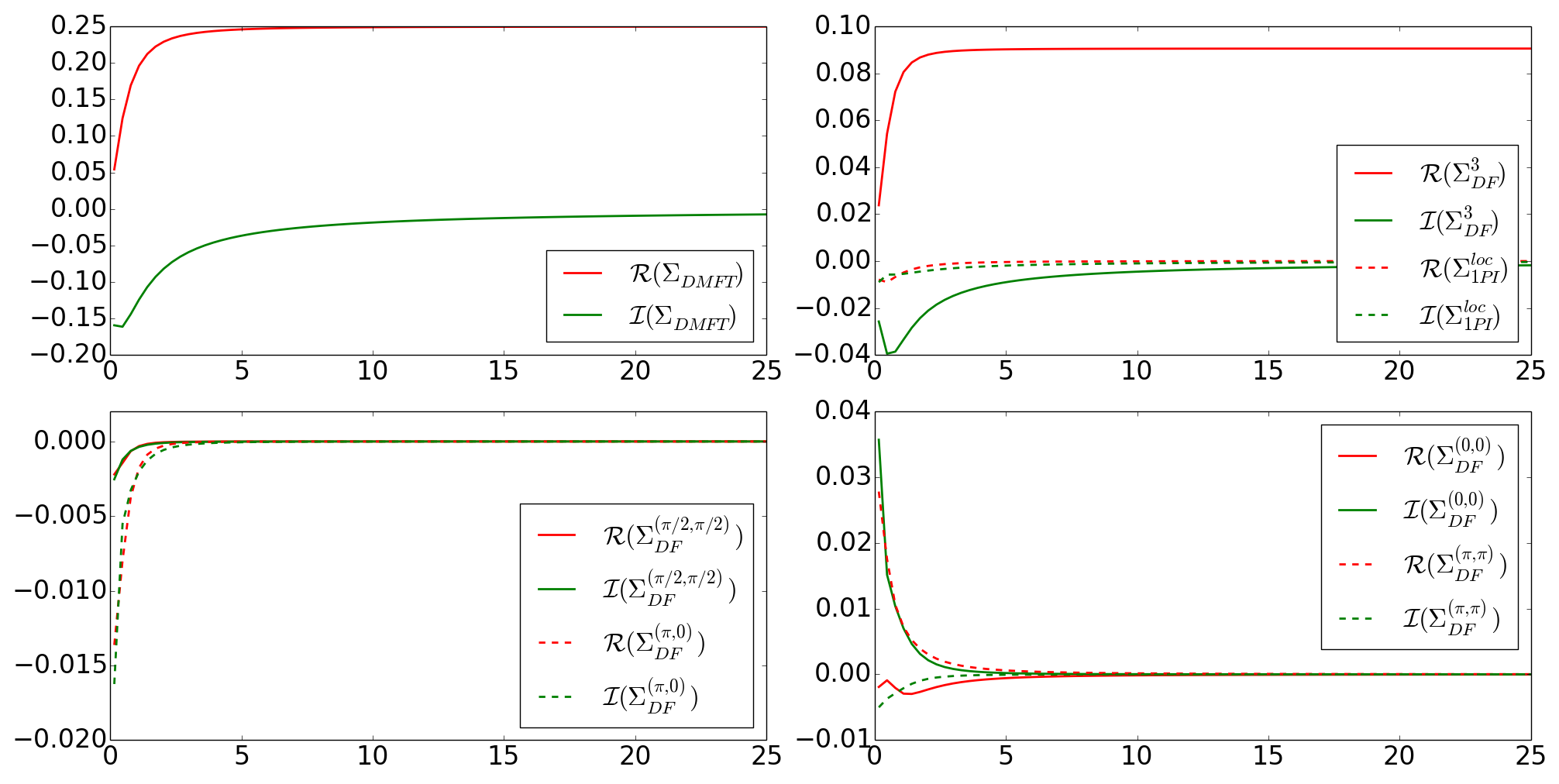}
\caption{(Color online)\label{3 corrs} Upper left: self-energy $\Sigma$ in DMFT for the square lattice FKM at quarter filling $p_1=0.25$, $U=1$, and $T=0.05$.
Upper right:  Contribution of the local diagrams based on the 3-particle vertex in Fig. \ref{3rdCorrDiag} (denoted as $\Sigma^3$) and the comparison to the local 1PI correction  ($\Sigma_{\rm 1PI}^{\rm loc}$) in a standard calculation based on the 2-particle vertex. Lower panels: Standard DF correction to the self-energy ($\Sigma_{DF}$) for four different ${\mathbf k}$-points as calculated from the two-particle vertex. Note that the correction term from the 3-particle vertex is of similar magnitude as the contributions from the conventional 2-particle vertex in DF; the DF self energies are those for the dual, not the real fermions.}
\end{minipage}
\end{figure*}

Fig.~\ref{3 corrs} presents different correction terms within 1PI and DF. The DMFT self-energy shows the usual behavior, with a Hartree-term proportional to $p_1 U$ as asymptotic value. The mathematical expressions for the correction to the DMFT self-energy within 1PI (upper right panel) and the dual self-energy within DF on the two-particle vertex level (lower panels) are very similar. In fact they only differ by the propagator used to close a loop in the respective diagram. Note, however, that the dual self-energy shown in Fig.~\ref{3 corrs}  has to be mapped back to extract the real self-energy. Hence, some care needs to be taken when comparing the two quantities. The additional contribution within 1PI leads to a local correction term for the self-energy on the level of two-particle vertices. Diagrammatically, 1PI is known to introduce corrections to the self-energy, that within DF stem from one-particle reducible contributions to the three-particle vertex, already from the two-particle vertex\cite{Katanin2013,1PI}.

Against this background, we compare in Fig.~\ref{3 corrs} (upper right panel)  the contributions based on the three-particle vertices within DF-theory to the local 1PI contribution calculated from the two-particle vertex, which already includes some of the contributions of the former.  The big difference between $\Re \Sigma_{\rm DF}^3$ and  $\Re \Sigma_{\rm 1PI}^{\rm loc}$ stems from the  renormalization of  the Hartree term in  third order DF. Apart from this Hartree contribution, the two contributions are comparable, albeit --of course-- they  differ quantitatively since the    third order DF includes further diagrams.

The conventional corrections to the dual self-energy stemming from the two-particle vertex
are shown in  Fig.~\ref{3 corrs} (lower panels). These corrections 
are ${\mathbf k}$-dependent, and  we selected   $4$ ${\mathbf k}$-points of high symmetry: $(0,0), (0,\pi), (\pi/2,\pi/2)$ and $(\pi,\pi)$. The corrections are of the same order of magnitude for all four ${\mathbf k}$-points. Most important for our estimate is that  the corrections due to the three-particle vertex $\Sigma_{\rm DF}^3$ are of the same order of magnitude as  $\Sigma_{\rm DF}$. Thus, a truncation of the DF approach on the two-particle level does not seem justified for this set of parameters.


\section{Conclusion}
\label{sec:conclusion}
We have derived analytical expressions for the local $n$-particle Green's 
function of the FKM within DMFT and from this, by removing all disconnected Feynman diagrams, for the full $n$-particle vertex. This main result of our paper can be found in Eq. (\ref{nVert}). While it was known before that the $n\!=\!3$-particle vertex vanishes at half-filling, we show that this is indeed the case for all odd $n$, whereas it is finite for even $n$.
Our analysis of the prefactor for the vertex reveals that it grows post exponentially, which is partially compensated by a fluctuating sign of this prefactor.

The calculated local vertex serves as a starting point for diagrammatic extensions of DMFT such as the DF, D$\Gamma$A and 1PI approach. The latter two require the fully irreducible and the one-particle irreducible vertex, respectively, whereas the DF approach is based on the full vertex which we calculated.
Hitherto, these approaches have been restricted to the $n\!=\!2$-particle vertex level which is an important approximation of these diagrammatic extensions. 

Our calculation of the local vertex for arbitrary $n$ now allows us to go beyond this level and estimate the impact of terms containing higher-order vertices. For general $n$, we find --at least for selected diagrams-- that  the series should be absolutely convergent despite the divergence of the prefactor of the vertex for $n\rightarrow \infty$.

In addition, we estimate the importance of the  three-particle vertex for
 the dual fermion self-energy by calculating numerically  the contribution of a selected diagram that includes the three-particle vertex (Fig. \ref{3rdCorrDiag}). We find that this three-particle vertex contribution
is  of the same magnitude as the standard DF self-energy calculated
from the two-particle vertex only.  This implies that the conventional truncation of dual-fermion expansions on the two-particle level can have a relevant impact on the numerical results. 
\\On the other hand, Gukelberger et al. \citep{DF2PDMC} found a good agreement between DF truncated at the two-particle-vertex level and diagrammatic determinant Monte Carlo for the Hubbard model with particle-hole symmetry. This is, at first sight, a contradiction. Let us note however that,
for the FKM, the three-particle vertex vanishes at  particle-hole symmetry.
This calls for a numerical calculation of the  three-particle vertex for the Hubbard model, which constitutes a challenging task.
 It might be feasible thanks to recent improvements of CTQMC calculations using improved estimators \citep{Gunacker16}, and might possibly allow for resolving  this apparent discrepancy.

\section{Acknowledgments}
\label{sec:ack}
Financial support is acknowledged from the  European Research Council under the European Union's Seventh Framework Program (FP/2007-2013) through ERC grant n.\ 306447 (TR, KH) and the joint Russian Science Foundation (RSF)/DFG Grant No. 16-42-01057 LI 1413/9-1 (G.R.). The plots were made using the matplotlib \cite{Matplotlib} plotting library for python.


\begin{thebibliography}{}
\bibitem{Metzner89a}
W.~Metzner and D.~Vollhardt, \newblock Phys. Rev. Lett. {\bf 62} 324 (1989).

\bibitem{MuellerHartmann89a}
E.~M{\"{u}}ller-Hartmann, \newblock Z. Phys. B {\bf 74} 507 (1989).

\bibitem{Georges92a}
A.~Georges and G.~Kotliar, \newblock Phys. Rev. B {\bf 45} 6479 (1992).

\bibitem{Jarrell92a}
M.~Jarrell, \newblock Phys. Rev. Lett. {\bf 69} 168 (1992).

\bibitem{Georges96a}
A.~Georges, G.~Kotliar, W.~Krauth and M.~Rozenberg, \newblock Rev. Mod. Phys. {\bf 68} 13 (1996).

\bibitem{dmft1} G. Kotliar, S. Y. Savrasov, K. Haule, V. S. Oudovenko, O. Parcollet, and C. A. Marianetti, Rev. Mod. Phys. {\bf 78}, 865 (2006).
\bibitem{dmft2} K. Held, Advances in Physics {\bf 56}, 829 (2007).

\bibitem{DCA}
M.~H. Hettler, A.~N. Tahvildar-Zadeh, M.~Jarrell, T. Pruschke, and  H. R. Krishnamurthy \newblock Phys. Rev. B {\bf 58} 7475 (R) (1998).

\bibitem{clusterDMFT}
G.~Kotliar, S.~Y. Savrasov, G.~P\'alsson and G.~Biroli, \newblock Phys. Rev. Lett. {\bf 87} 186401 (2001).

\bibitem{LichtensteinDCA}
A.~I. Lichtenstein and M.~I. Katsnelson, \newblock Phys. Rev. B {\bf 62} 9283 (R) (2000).

\bibitem{Maier04}
T.~Maier, M.~Jarrell, T.~Pruschke and M.~H. Hettler, \newblock Rev. Mod. Phys. {\bf 77} 1027 (2005).


\bibitem{DGA1} A.~Toschi, A.~A. Katanin, and K.~Held, {Phys. Rev. B} \textbf{%
75}, 045118 (2007); {Prog. Theor. Phys. Suppl.} \textbf{176}, 117 (2008).

\bibitem{Kusunose06} H. Kusunose, {J. Phys. Soc. Jpn}  {\bf 75},  {054713} (2006.
 
\bibitem{Katanin09}  A.\ A. Katanin, {A. Toschi}, and K. Held, Phys. Rev. B
\textbf{80}, 075104 (2009).

\bibitem{DualFermion} A. N. Rubtsov, M. I. Katsnelson, and A. I. Lichtenstein, Phys. Rev. B \textbf{77}, 033101 (2008).

\bibitem{DualBoson} A. N. Rubtsov, M. I. Katsnelson, and A. I. Lichtenstein, Annals of Physics \textbf{327}, 1320 (2012).

\bibitem{DGAintro} K. Held, "Dynamical vertex approximation", in E. Pavarini, E. Koch, D. Vollhardt, A. Lichtenstein (Eds.): "Autumn School on Correlated Electrons. DMFT at 25: Infinite Dimensions", Reihe Modeling and Simulation, Vol. 4 (Forschungszentrum Julich, 2014) [arXiv:1411.5191].


\bibitem{AbinitioDGA}
A. Toschi, G. Rohringer, A. A. Katanin, and K. Held, Annalen der Physik, {\bf 523},  698 (2011). 

\bibitem{Li15}G. Li,  Phys.Rev. B {\bf 91}, 165134 (2015).

\bibitem{1PI}G. Rohringer, A. Toschi, H. Hafermann, K. Held, V.I. Anisimov, A. A. Katanin, Phys. Rev. B {\bf 88}, 115112 (2013).
 
\bibitem{DMF2RG}
C. Taranto, S. Andergassen, J. Bauer, K. Held, A. Katanin, W. Metzner, G. Rohringer, A. Toschi, Phys. Rev. Lett. {\bf 112}, 196402 (2014).

\bibitem{Ayral15}
T. Ayral and O. Parcollet {Phys Rev. B} {\bf 92}, {115109} (2015).

\bibitem{Ayral16} T. Ayral and O. Parcollet, Phys Rev. B {\bf 94}, 075159 (2016).

\bibitem{Kitatani15} M. 
  {Kitatani}, N. Tsuji, and H. Aoki, {Phys. Rev. B} {\bf 92}, {085104} (2015).

\bibitem{note_fourier} Here we have used the symmetric Fourier transform between the Matsubara time and the Matsubara frequency space, i.e., $(1/\sqrt{\beta})\int_0^{\beta}d\tau\ldots$ for the Fourier transform and $(1/\sqrt{\beta})\sum\nu\ldots$ for the inverse Fourier transform. This convention is convenient for the decomposition of the $n$-particle Green's function into its connected and unconnected parts in the FKM as performed in the following, since it avoids any prefactors $\beta^n$ in the definition of $G^{(n)}$ in frequency space. 

\bibitem{note_prefactor_diagram} Note that the sum should in principle run over $\nu_1\ldots\nu_{2n}$ since the $n$-particle Green's function $G^{(n)}$ (and, hence, also $G^{(1)}$) depends on frequencies $\nu_1\ldots \nu_n$ (i.e., on $\nu_1$ and $\nu_2$ for $G^{(1)}$). Considering the convention we used for the Fourier transform in [\onlinecite{note_fourier}] this yields a prefactor $1/\sqrt{\beta}^{2n}=1/\beta^n$. Moreover, in Eq.~(\ref{equ:gendiagram}) we have already performed the sums over $\nu_{n+1}\ldots\nu_{2n}$ which just eliminates the corresponding $\delta$-functions in the definition of the Green's function.

\bibitem{Rohringer11}
G. Rohringer, A. Toschi, A. A. Katanin, and K. Held, Phys. Rev. Lett. {\bf 107}, 256402 (2011) 

\bibitem{Hirschmeier15} 
D. Hirschmeier, H. Hafermann, E. Gull, A. I. Lichtenstein,
and A. E. Antipov, Phys. Rev. B.
{\bf 92}, 144409 (2015).

\bibitem{Schaefer17}
    T. Sch\"afer, A. A. Katanin, K. Held, and A. Toschi,  arXiv:1605.06355.

\bibitem{Antipov14} A. E. Antipov, E. Gull, and S. Kirchner,
 Phys. Rev. Lett. {\bf 112}, 226401 (2014).

\bibitem{Schaefer15}
T. Sch\"afer, F. Geles, D. Rost, G. Rohringer, E. Arrigoni, K. Held, N. Bl\"umer, M. Aichhorn, and A. Toschi, Phys. Rev. B {\bf 91}, 125109 (2015).
 
\bibitem{footnote1}
Besides, there is also the question of which diagrams to generate with the local two-particle vertex as a starting point.  Here, in principle, the parquet equations\cite{parquet,Bickers04} or the diagrammatic Mote Carlo approach\cite{Iskakov16} are the most extensive since they generate all Feynman diagrams for a given fully irreducible vertex. But in practice more often simpler ladder diagrams are summed up, with a few noteworthy exceptions solving the parquet equations.\cite{Yang09,Tam13,Valli15,Li16,Janis16}

\bibitem{parquet} C. De Dominicis, J. Math. Phys. {\bf 3}, 983 (1962);  C. De Dominicis and P. C. Martin,{\bf 5}, 14 (1964).

\bibitem{Bickers04} 
J. Bickers, Chap 6 in {\it Theoretical Methods for Strongly C
orrelated Electrons},  eds. D. Senechal, A.M. Tremblay, and C. Bourbonnais (Springer, New York, 2004).

\bibitem{Iskakov16}
S. Iskakov, A. E. Antipov, and E. Gull,
Phys. Rev. B {\bf 94}, 035102 (2016).

\bibitem{Yang09}
 S. X. Yang, H. Fotso, J. Liu, T. A. Maier, K. Tomko, E. F.
 D'Azevedo, R. T. Scalettar, T. Pruschke, and M. Jarrell, Phys. Rev. E {\bf 80},
 046706 (2009).

\bibitem{Tam13} Ka-Ming Tam, H. Fotso, S.-X. Yang, Tae-Woo Lee, J. Moreno, J. Ramanujam, and M. Jarrell,  Phys. Rev. E {\bf 87}, 013311 (2013).

\bibitem{Valli15}
A. Valli, T. Sch\"afer, P. Thunstr\"om, G. Rohringer, S. Andergassen, G. Sangiovanni, K. Held, and A. Toschi, Phys. Rev. B {\bf 91}, 115115 (2015).


\bibitem{Li16}
G. Li, N. Wentzell, P. Pudleiner, P. Thunstr\"om, and K. Held, Phys. Rev. B 93, 165103 (2016).

\bibitem{Janis16}
    V. Jani\v{s}, A. Kauch, and V. Pokorn\'y,  arXiv:1604.01678.

\bibitem{vertex}
G. Rohringer, A. Valli, and A. Toschi, Phys. Rev. B {\bf 86}, 125114 (2012).




\bibitem{Rubtsov04}
A. Rubtsov and A. Lichtenstein, Journal of Experimental
and Theoretical Physics Letters
{\bf 80}, 61 (2004)

\bibitem{Werner06}
P.   Werner,   A.   Comanac,   L.   de'   Medici,   M.   Troyer,
and  A.  J.  Millis,  Phys.  Rev.  Lett.
{\bf 97},  076405  (2006).

\bibitem{Gull11}
E.  Gull,    A.   J.   Millis,    A.   I.   Lichtenstein,    A.   N.
Rubtsov,  M.  Troyer,  and  P.  Werner,  Rev. Mod. Phys.
{\bf 83}, 349 (2011).
\bibitem{Gunacker15}
P.  Gunacker,   M.  Wallerberger,   E.  Gull,   A.  Hausoel,
G.   Sangiovanni,    and   K.   Held,    Phys.   Rev.   B
{\bf 92},
155102 (2015).

\bibitem{Gunacker16}
    P. Gunacker, M. Wallerberger, T. Ribic, A. Hausoel, G. Sangiovanni, and K. Held,   arXiv:1607.01211.

\bibitem{Falicov69} L. M. Falicov and J. C. Kimball, Phys. Rev. Lett. {\bf 22}, 997 (1969).



\bibitem{Brandt89}
U. Brandt and C. Mielsch, Z. Phys. B: Condens. Matter {\bf 75}, 365 (1989).


\bibitem{Freericks03}
J. K. Freericks and V. Zlati\'c, Rev. Mod. Phys. {\bf 75}, 1333 (2003).

\bibitem{vanDongen90}
P.G.J. van Dongen and D. Vollhardt, Phys. Rev. Lett. {\bf 65}, 1663
(1990).

\bibitem{Freericks00}
 J. K. Freericks and P. Miller,  Phys. Rev. B {\bf 62}, 10022 (2000).


\bibitem{Janis10}
V. Jani\v{s} and V. Pokorn\'y, Phys. Rev. B {\bf 81}, 165103 (2010).

\bibitem{Pokorny13}
V. Pokorn\'y and V. Jani\v{s}, J. Phys.: Condens. Matter {\bf 25}, 175502 (2013).


\bibitem{Antipov16} A. E. Antipov, Y. Javanmard, P. Ribeiro and S. Kirchner Phys. Rev. Lett. {\bf 117} 146601  (2016). 

\bibitem{Ribic16} T. Ribic, G. Rohringer and K. Held,  Phys. Rev. B {\bf 93} 195105 (2016). 

\bibitem{Shvaika}
A. M. Shvaika, Physica C {\bf 341}, 177 (2000).

\bibitem{DF2PDMC} J. Gukelberger, E. Kozik, and H. Hafermann,
arxiv:1611.07523.

\bibitem{Matplotlib} J. D. Hunter,
Computing In Science \& Engineering {\bf 9}(3):90--95 (2007).

\bibitem{note_1PD} Note that there are of course diagrams which are 1PD in more than one frequency. In fact, an unconnected diagram consisting of a product of $n$ one-particle Green's functions belongs also to this class and is 1PD in all its frequency arguments.

\bibitem{halffilling} Note that for half-filling and $\mathcal{C}_n$ vanishes for all odd numbers $n$. This can be readily proved by induction considering the vanishing of $\mathcal{F}_{n}$ for all odd $n$ [see Eq.~(\ref{Fl})].


\bibitem{RohringerPhD} G. Rohringer, Ph.D. Thesis, TU Wien (2013).

\bibitem{RMP}  G. Rohringer, H. Hafermann, A. Toschi, A.A. Katanin, A.E. Antipov, M.I. Katsnelson, A.I. Lichtenstein, A.N. Rubtsov, and K. Held, Rev. Mod. Phys. (commissioned).
\bibitem{Katanin2013} A A Katanin,  J. Phys. A: Math. Theor. {\bf 46}, 045002 (2013).
\end{thebibliography}
\end{document}